\documentclass[aps,prx,twocolumn,longbibliography,floatfix,10pt,superscriptaddress]{revtex4-2}
\usepackage[utf8]{inputenc}
\usepackage{mathrsfs}
\usepackage{natbib}
\usepackage{graphicx}
\usepackage{latexsym}
\usepackage{amssymb}
\usepackage{amsmath}
\usepackage{amsfonts}
\usepackage{gensymb}
\usepackage{bm}
\usepackage{hyperref}
\usepackage{braket}
\usepackage{physics}
\usepackage{bbold}
\usepackage[caption=false]{subfig}
\usepackage[dvipsnames]{xcolor}
\usepackage[normalem]{ulem}
\usepackage{siunitx}
\usepackage{mhchem}
\usepackage{multirow}
\usepackage{soul}

\usepackage{slashed}

\definecolor{myforestgreen}{RGB}{34,139,34}

\newcommand{\supplementarysection}{%
 \setcounter{figure}{0}
 \let\oldthefigure\thefigure
 \renewcommand{\thefigure}{S\oldthefigure}
 \setcounter{section}{0}
 \setcounter{equation}{0}
 \let\oldtheequation\theequation
 \setcounter{table}{0}
 \let\oldthetable\thetable
 \renewcommand{\thetable}{S\oldthetable}
}

\newenvironment{dfn}{{\vspace*{1ex} \noindent \bf Definition }}{\vspace*{1ex}}

	\newcommand{\beq}{\begin{eqnarray}}
	\newcommand{\eeq}{\end{eqnarray}}
	\newcommand{\bea}{\begin{eqnarray}\begin{aligned}}
	\newcommand{\eea}{\end{aligned}\end{eqnarray}}

	\renewcommand{\v}{{\bf v}}

\begin{document}

\title{Chiral superconductivity near a fractional Chern insulator}

\author{Taige Wang}
\affiliation{Department of Physics, University of California, Berkeley, CA 94720, USA \looseness=-2}

\author{Michael P. Zaletel}
\affiliation{Department of Physics, University of California, Berkeley, CA 94720, USA \looseness=-2}

\begin{abstract}

Superconductivity arising from fully spin-polarized, repulsively interacting electrons can host intrinsically chiral Cooper pairs and Majorana zero modes, yet no concrete microscopic route to such a state has been established. Motivated by recent observations in twisted homobilayer MoTe$_2$ and rhombohedral pentalayer graphene, where fractional Chern insulators (FCIs) appear adjacent to spin-valley-polarized superconductors, we investigate a minimal model: spinless electrons in the lowest Landau level subject to a tunable periodic potential. Large-scale density-matrix renormalization group (DMRG) calculations reveal that, as the FCI gap closes, two nearly degenerate phases emerge before the system turns metallic: a chiral \(f\)-wave superconductor and a \(\sqrt{3} \times \sqrt{3}\) charge-density wave (CDW) whose energies differ by less than \(1\%\). These two competing states mirror the superconducting and re-entrant integer quantum Hall (RIQH) phases observed experimentally near the FCI regime. The superconducting dome survives realistic Coulomb interaction, light doping, and various lattice geometry. Melting the FCI therefore provides a new mechanism for realizing spin-polarized chiral superconductivity and RIQH order. We predict that twisted MoTe$_2$ at larger twist angles will develop a superconducting dome even at filling \(\nu = 2/3\), and suppressing this superconductivity with a magnetic field should drive the system into an RIQH state.

\end{abstract}

\maketitle

\begin{figure}[t]
 \centering
	\includegraphics[width=\columnwidth]{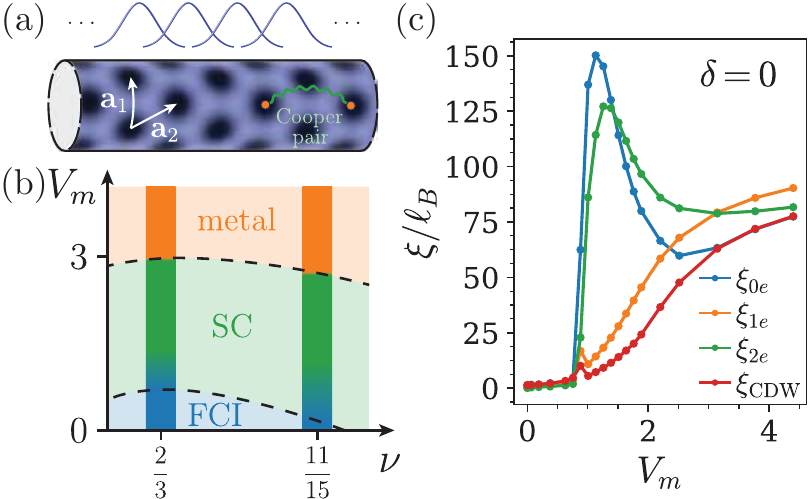}
	\caption{\textbf{Triangular-lattice LLL model and resulting phase diagram.}
(a) Infinite cylinder with spin-polarized LLL orbitals (schematic curves) and triangular lattice potential (color scale). Dark color denotes low potential energy and therefore high electron density. Vectors $\mathbf a_{1,2}$ define the YC lattice wrapping the cylinder. 
(b) Numerical phase diagram at $L_y=5\ell_B$ and DMRG bond dimension $\chi=10000$. A fractional Chern insulator (FCI, blue) and a compressible metal (orange) are separated by a dome of chiral superconductor (green). Dashed lines interpolate phase boundaries obtained at $\nu=2/3$ and $11/15$. 
(c) Correlation lengths at $\nu=2/3$ for the same cylinder and $\chi$: neutral $\xi_{0e}$ (blue), single-electron $\xi_{1e}$ (orange), Cooper-pair $\xi_{2e}$ (green), and density-wave $\xi_{\mathrm{CDW}}$ (purple) sectors are shown. The window $\xi_{2e}>\xi_{1e}$ signals Luther–Emery liquid behavior, while $\xi_{2e}>\xi_{0e}$ marks the quasi-1D superconducting regime.}
\label{fig:pd}
\end{figure}

\section{Introduction}

In conventional metals phonons can bind electrons into singlet Cooper pairs~\cite{Bardeen1957}; in various correlated materials magnetic fluctuations have been argued to play the same role ~\cite{moriya2003antiferromagnetic,RossatMignod1991,Scalapino1995}.
Removing one spin species eliminates both glues and raises a fundamental question: \emph{Can a purely repulsive, fully spin-polarized electron fluid still form a superconductor}? 
An affirmative answer would be significant, because any such condensate is expected to be \emph{chiral} and might host Majorana modes~\cite{volovik1999fermion, vollhardt2013superfluid, ReadGreen2000,Nayak2008}.

Two-dimensional moiré materials have placed this puzzle on experimental footing. 
Twisted MoTe$_2$ homobilayers and rhombohedral pentalayer graphene realize ultra-flat, spin-polarized Chern bands. At filling $|\nu|=2/3$ they stabilize a fractional Chern insulator (FCI)~\cite{Cai2023,Park2023,Zeng2023,Xu2023,Lu2024}. 
Yet modest doping in MoTe$_2$ or weakening the moiré in rhombohedral graphene collapses the FCI and unveils a spin-valley-polarized superconductor—despite the absence of any obvious pairing glue~\cite{Xu2025FCISC,Han2025}. 
Weak-coupling theories hint at superconducting instabilities, but their assumptions need not apply in the strongly correlated regime that hosts FCI~\cite{Geier2024,Chou2025,ParraMartinez2025,Wang2024,ShavitAlicea2025,Xu2025,guerci2025,JahinLin2025,maymann2025how,dong2025controllable,JahinLin2025}.

Here we propose a minimal model that explains how an FCI can melt into \emph{either} of two competing descendants—a spin-polarized chiral superconductor (SC) or a re-entrant integer quantum Hall (RIQH) phase. Our minimal model treats spinless electrons confined to the lowest Landau level and subject to a tunable periodic potential, capturing the key ingredients of twisted MoTe$_2$ and related moiré systems: a single, nearly ``ideal Chern band'' with adjustable bandwidth~\cite{Ledwith2022Ideal,Ledwith2023Vortexable,Yu2025QuantumGeometry,Dong2023CFL,Dong2024Pentalayer,Yu2024,HerzogArbeitman2024}. 
For spin-polarized Chern bands that support FCIs, this model plays a role analogous to the one-band Hubbard model for cuprates.

In this work we conduct large-scale infinite-cylinder density–matrix renormalization group (DMRG) calculations and find non-perturbative evidence that a purely repulsive, fully spin-polarized fluid can condense into a superconducting state. 
Starting from a $\nu = 2/3$ FCI in the flat-band limit, as the bandwidth increases we find the expected FCI–metal transition is pre-empted by two nearly degenerate phases whose energies differ by less than \(1\%\): 
(i) a dome of chiral \(f\)-wave superconductor that evolves smoothly from tightly bound, BEC-like pairs on the FCI side to weak-coupling, BCS-like behavior on the metallic side, and 
(ii) a commensurate \(\sqrt3\times\sqrt3\) charge-density wave carrying Hall conductance \(\sigma_{xy}=e^2/h\), mirroring the re-entrant integer quantum Hall (RIQH) state observed experimentally by doping the FCI in both twisted MoTe$_2$ and rhombohedral pentalayer graphene~\cite{Xu2025FCISC,Lu2024,LuHanYao2025}. Thus melting an FCI provides a unified microscopic mechanism for both spin-polarized chiral superconductivity and RIQH order in moiré Chern bands.

We outline a field-theoretic framework in which collapsing a FCI can give rise to \emph{either} a chiral superconductor or a RIQH state, without invoking dopant-induced anyon superconductivity~\cite{Darius2024}. 
Regardless of its microscopic mechanism, the pairing we find shows that the FCI-to-metal boundary is a promising place to look for spin-polarized superconductivity in Chern bands and gives us a simple model for studying this regime further. Two immediate experimental tests follow from our study. First, moving to larger twist angles beyond the narrow-band FCI limit should broaden the superconducting dome to encompass filling factor $\nu = 2/3$. Second, applying a magnetic field strong enough to quench the superconductivity should expose the closely competing RIQH phase, as already observed in twisted MoTe$_2$~\cite{Xu2025FCISC}.

\section{Triangular-lattice LLL model}

We study spin-polarized electrons confined to the lowest Landau level (LLL) on an infinite cylinder of circumference $L_y$ as shown in Fig.~\ref{fig:pd}(a). 
The electrons interact via a two-body interaction $V(\mathbf q)$ and experience a triangular lattice potential of amplitude $V_m$,
\begin{equation}
\hat H= \frac12 \sum_{\mathbf q}V(\mathbf q) \rho_{-\mathbf q}\rho_{\mathbf q}
    - \frac{V_m}2\sum_{j}\rho_{\mathbf G_j},
\end{equation}
where $\rho_{\mathbf q}$ is the bare density operator and $\{\mathbf G_j\}$ are the six shortest reciprocal-lattice vectors. 
The lattice constant $a$ of the triangular lattice is chosen so that there is one flux quantum per-unit cell $\frac{\sqrt{3}}{2} a^2 = 2 \pi \ell_B^2$, 
resulting in a single $C=1$ Chern band without Hofstadter sub-bands. The lowest Landau level (LLL) projected single-particle dispersion is 
\begin{equation}
  \epsilon_{\mathbf k} = V_m e^{-\frac{\pi}{\sqrt{3}}} \sum_{j=1}^3 \cos ( \mathbf k \cdot \mathbf a_j )
\end{equation}
where $\mathbf a_{1, 2, 3}$ are the Braivas vectors shown in Fig.~\ref{fig:pd}(a) with $\mathbf a_{3} = \mathbf a_2 - \mathbf a_1$.
This model preserves the triangular symmetry of twisted MoTe$_2$ and rhombohedral pentalayer graphene while imposing an \emph{ideal} band geometry, i.e., strictly uniform Berry curvature and a vanishing trace condition~\cite{Ledwith2022Ideal,Ledwith2023Vortexable,Yu2025QuantumGeometry}. 
In the actual materials these quantum-geometric quantities deviate rather weakly from the ideal limit~\cite{Dong2023CFL,Dong2024Pentalayer,Yu2024,HerzogArbeitman2024}, so our model captures the essential physics.
We note our model has in addition a $C_2$ symmetry, which these valley-polarized materials do not.

After projection to the LLL, ground states are obtained with infinite-cylinder DMRG that conserves the global $U(1)$ charge and the momentum $k_y$~\cite{ExactMPS,TopoChara}. 
Unless stated otherwise we use a pure contact repulsion $V(\mathbf r)= 4 \pi \ell_B^2
\nabla^2 \delta(\mathbf{r})$ (Haldane pseudopotential $V_1 = 1$)~\cite{Haldane1983} and impose ``YC'' boundary conditions, i.e., one in which triangular Bravais vector $\mathbf a_1 = (0,a)$ wraps around the cylinder.
Circumferences $L_y=4a-9a$ are explored. The effects of a gate-screened Coulomb tail and of a square-lattice potential are analyzed at the end.

\section{Superconducting dome}

We concentrate on fillings in the vicinity of the $\nu=2/3$ Jain state and define the doping $\delta\equiv\nu-2/3$; fillings near $\nu=1/3$ yield an analogous phase diagram (see Supplementary Material). 
Fig.~\ref{fig:pd}(b) presents the numerical phase diagram as a function of lattice amplitude $V_m$ for a cylinder of circumference $L_y = 5\ell_B$. 
Phase boundaries are determined from the correlation lengths of four quantum number sectors: the neutral sector $\xi_{0e}$, the single-electron sector $\xi_{1e}$, the Cooper-pair sector $\xi_{2e}$, and the density-wave sector $\xi_{\mathrm{CDW}}$ (Fig.~\ref{fig:pd}(c)). Here, $\xi_{1e}$ and $\xi_{2e}$ denote the correlation lengths in sectors carrying charges $e$ and $2e$, respectively, while $\xi_{0e}$ and $\xi_{\mathrm{CDW}}$ both correspond to charge-neutral sectors—the former at zero momentum $k_y$, and the latter at finite $k_y$.

At $\delta=0$ the two extremes of $V_m$ behave as expected. 
For $V_m=0$ the model reduces to the parent $\nu=2/3$ Hamiltonian~\cite{Haldane1983}; all correlation lengths remain short and the entanglement spectrum exhibits the anti-chiral boson counting, confirming a gapped fractional Chern insulator (FCI, blue). 
At large $V_m$ the kinetic energy dominates, a Fermi surface emerges (Fig.~\ref{fig:pair}(b)), $\xi_{1e}$ surpasses every other correlation length, and the ground state is a compressible metal (orange)~\footnote{In the metal, $\xi_{1e}$ scales algebraically with DMRG bond-dimension $\chi$.}.

\begin{figure}[htb]
 \centering
  \includegraphics[width=\columnwidth]{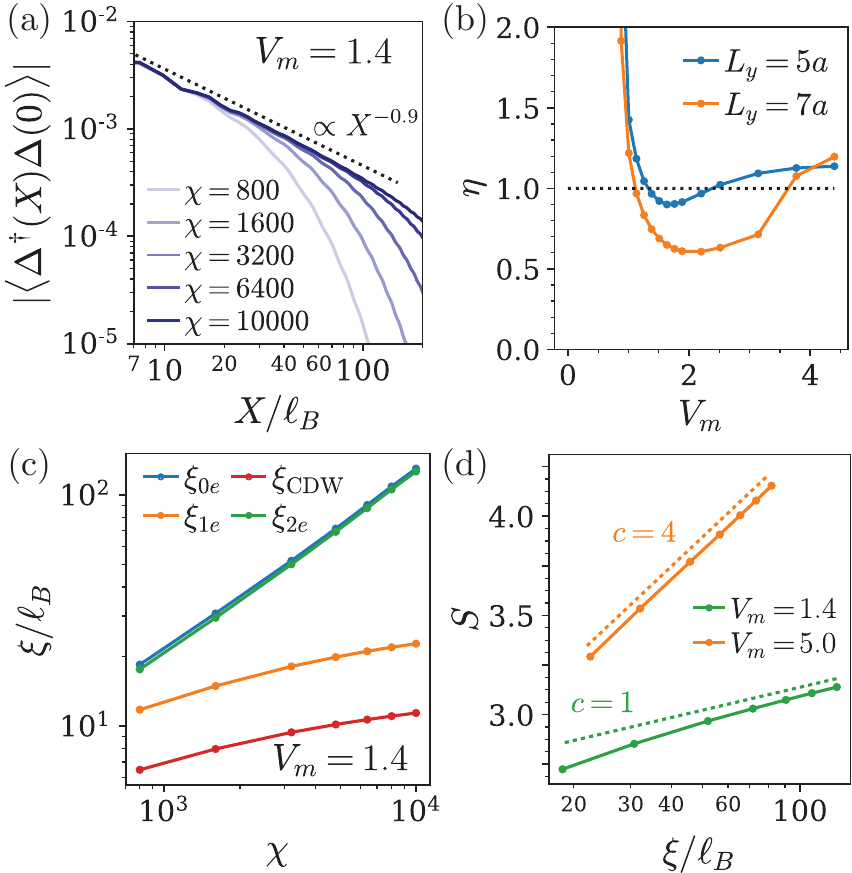}
	\caption{\textbf{Luther–Emery liquid inside the superconducting dome at $\boldsymbol{\nu=2/3}$.}
(a) Pair correlator $|\langle\Delta^\dagger(X)\Delta(0)\rangle|$ at $V_m = 1.4$ and $L_y = 5a$ for increasing bond dimension $\chi$. 
A power-law decay $X^{-\eta}$ with $\eta \simeq 0.9$ confirms algebraic long-range order of the Cooper pairs. 
(b) Luttinger exponent $\eta$ versus lattice amplitude $V_m$ for $L_y = 5 a$ (blue) and $L_y = 7a$ (orange). 
(c) Correlation lengths at $V_m = 1.4$ and $L_y = 5a$ as a function of $\chi$. 
(d) Entanglement entropy $S$ versus correlation length $\xi$ tuned by $\chi$. 
A log-linear fit gives central charge $c \approx 1$ in the superconducting dome ($V_m = 1.4$, green) and $c \approx 4$ in the metallic phase ($V_m = 5.0$, orange).}
\label{fig:corr}
\end{figure}

The most striking feature is a broad \emph{superconducting dome} that wedges itself between the topologically ordered FCI and the compressible metal. 
As soon as the topological gap closes, the Cooper-pair correlation length overtakes the single-electron length $\xi_{2e}>\xi_{1e}$, signaling a spinless analog of the Luther–Emery (LE) liquid in which single electrons are gapped while pair correlations decay algebraically~\cite{LutherEmery1974,Emery1976}. 
At the dome center $\xi_{2e}$ reaches nearly ten times of the cylinder circumference and, within a narrower window, even exceeds the neutral correlation length $\xi_{2e}>\xi_{0e}$, marking a quasi-1D superconducting regime (green band in Fig.~\ref{fig:pd}(b)). 
The finite $\xi_{2e}$ is an artifact of the finite bond dimension $\chi$ used in our simulations; in Fig.~\ref{fig:corr}(c) we show that $\xi_{2e}$ diverges algebraically with $\chi$, in contrast to $\xi_{1e}$ which saturates.
On wider cylinder $L_y=7a$, $\xi_{2e}>\xi_{0e}$ in most of the LE phase (see Supplementary Material), suggesting that pairing correlations will only grow as the system approaches two dimensions.

Electron doping to $\delta=1/15$ ($\nu=11/15$) preserves the same three-phase sequence: an FCI at small $V_m$, a superconducting dome at intermediate $V_m$, and a metal at large $V_m$. 
Within the doped FCI the density-wave channel correlation length becomes longest $\xi_{\mathrm{CDW}}>\xi_{0e}$, indicating an incipient anyon Wigner crystal and pushing the transition out of FCI to lower $V_m$ (see Supplementary Material). 
Pairing is enhanced throughout compared to at $\nu = 2/3$: $\xi_{2e}$ exceeds its undoped value across the dome. 
The hole-doped case $\delta=-1/15$ is omitted because it realizes a distinct $\nu=3/5$ Jain state.

The close similarity between the $\delta = 0$ and $\delta = 1/15$ phase diagrams indicates that superconductivity originates from melting the parent FCI itself, rather than from dopant-induced Hall states in the ``anyon superconductivity'' scenario~\cite{Laughlin1988PRL,Laughlin1988Science,HalperinAnyonSC1992,FisherLeeAnyonSC1991,Darius2024}. From now on, we will focus on filling $\nu = 2/3$ till the end when we discuss the effect of square lattice potential.

\section{Luther–Emery liquid}

On an infinite cylinder the system is effectively one-dimensional, so a continuous $U(1)$ symmetry cannot break and the Cooper pair field satisfies $\langle\Delta\rangle = 0$. 
Superconductivity must therefore manifest as a Luther–Emery (LE) liquid~\cite{LutherEmery1974,Emery1976,GannotKivelson2023}: single electrons are gapped, whereas Cooper pairs remain gapless. 
Consequently the single-electron Green’s function decays exponentially $|\langle c^{\dagger}(X)c(0)\rangle| \propto e^{-X/\xi_{1e}}$, while the pair correlator shows algebraic order $|\langle\Delta^{\dagger}(X)\Delta(0)\rangle| \propto X^{-\eta}$. 
The Luttinger parameter $K = 1/\eta$ sets the dominant tendency: $\eta<1$ favors superconductivity, $\eta>1$ favors charge density wave~\cite{jiang2019superconductivity,troyer1993spin}.

To investigate this decay numerically we note that the specific real-space profile of the pair field \(\Delta\) on the cylinder does not influence the asymptotic scaling of its correlator as long as \(X \gg L_y\). We therefore adopt a definition that is convenient for our matrix produce state (MPS) ansatz. In Landau gauge the single-particle orbitals are labeled by their transverse momentum \(k_y = 2\pi j/L_y\) with \(j \in \mathbb{Z}\). For Fig.~\ref{fig:corr}(a) we define the pair operator
\begin{equation}
  \Delta(X) = c^{\dagger}_{i+X}\,c^{\dagger}_{j+X},
\end{equation}
with \((i,j)\) specifying the pair and the separation \(X\) taken in integer multiples of \(L_y/a\). The correlations peak whenever \(i + j \equiv 0 \pmod{L_y/a}\), indicating zero-momentum pairing (\(k_y = 0\)). Fig.~\ref{fig:corr}(a) picks the strongest one among zero-momentum channels. 
Throughout the superconducting dome the pair correlator follows a power law over nearly two decades in distance, while the single-electron correlator decays exponentially (see Supplementary Material), confirming gapped quasiparticles but gapless Cooper pairs. 
A finite interval in $V_m$ satisfies $\eta<1$ as shown in Fig.~\ref{fig:corr}(b). This interval widens and $\eta$ decreases when the circumference is increased from $L_y=5a$ to $7a$, indicating that pairing correlations strengthen as the system approaches two dimensions.

\begin{figure}[t]
 \centering
  \includegraphics[width=\columnwidth]{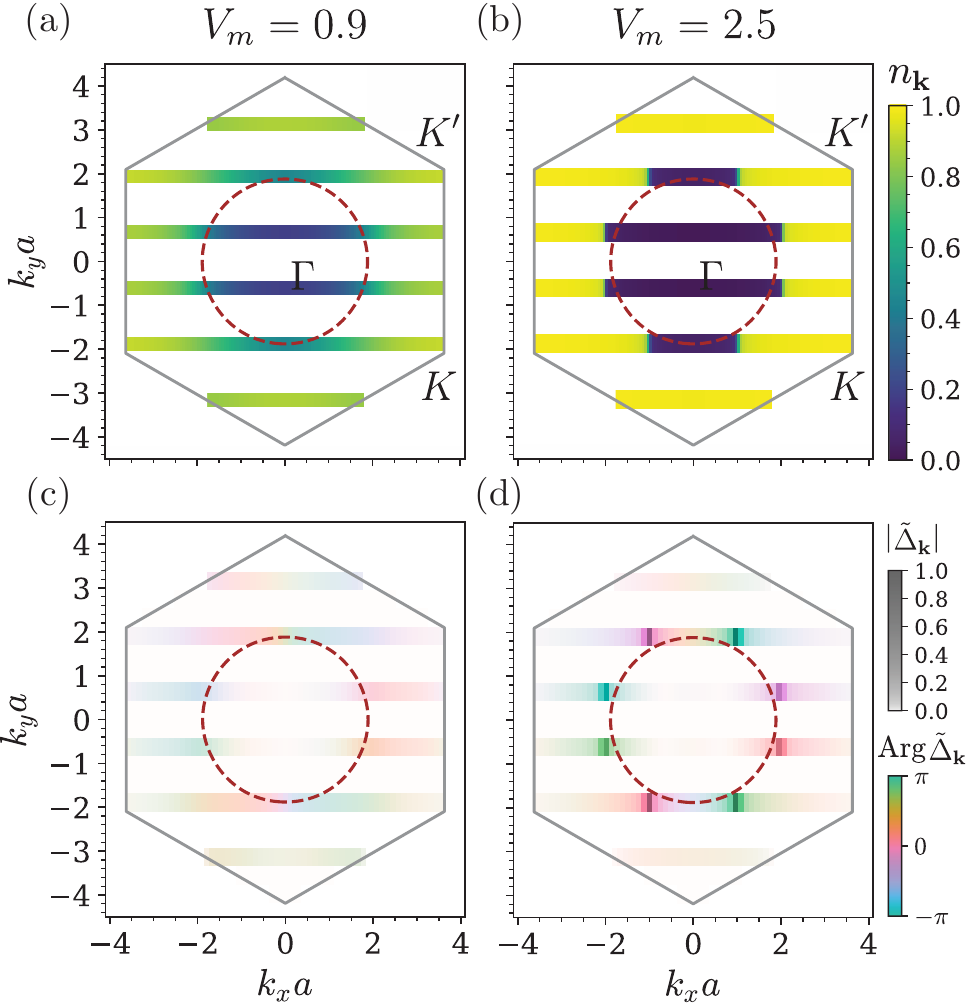}
\caption{\textbf{Momentum-space structure of the chiral $f$-wave pairing state.}
(a,b) Single-particle occupation $n_{\mathbf k}$ in the Brillouin zone (grey hexagon) for lattice amplitudes $V_m=0.9$ (left, FCI side of the dome) and $V_m=2.5$ (right, metallic side). Horizontal strips show the allowed $k_y$ values on the $L_y=5a$ cylinder; red dashed contour marks the non-interacting Fermi surface. Steps in $n_{\mathbf k}$ reveals the Fermi surface.
(c,d) Magnitude (greyscale) and phase (hue) of the proxy gap function $\tilde{\Delta}_{\mathbf k}$ for the same two values of $V_m$. In both cases the phase winds by $-6\pi$ around the Fermi pocket, characteristic of chiral $f - if$ pairing ($\mathcal{C}_{\mathrm{BdG}}=-1$ due to parent Chern band). 
}
\label{fig:pair}
\end{figure}

Correlation lengths and the central charge provide additional evidence for the LE liquid phase. As the DMRG bond dimension $\chi$ grows, $\xi_{1e}$ appears to saturate while $\xi_{2e}\propto\chi^{\kappa}$ diverges as a power-law in $\chi$, reassuring gapped quasiparticles but gapless Cooper pairs (Fig.~\ref{fig:corr}(c)). Furthermore, fits of the entanglement entropy $S=(c/6)\ln\xi+\text{const}$ yield a central charge $c \simeq 1$ inside the dome (Fig.~\ref{fig:corr}(d))—consistent with a single gapless Goldstone mode of a superconductor.
Outside the dome, where the Fermi surface intersects four momentum wires at $L_y = 5a$ (Fig.~\ref{fig:pair}(b)), the same analysis gives $c \simeq 4$ as expected.

The simultaneous observation of algebraic pair order, divergent $\xi_{2e}$ alongside a saturated $\xi_{1e}$, and a central charge $c \simeq 1$ unequivocally identifies the entire superconducting dome as a LE liquid phase on the $L_y = 5a$ and $7a$ cylinder. The decrease in $\eta$ with system size suggests that the superconducting dome will evolve smoothly into a superconductor in the two-dimensional limit.

\section{Pairing symmetry and the BEC–BCS crossover}

Identifying the pairing symmetry is crucial for any unconventional superconductor. It can be probed by phase-sensitive experiments, benchmarked against analytic theory, and dictates the BdG Chern number and therefore the count of chiral Majorana modes. Because $\langle\Delta\rangle=0$ on the cylinder, we turn to the leading long-range charge-$2e$ fluctuation. Diagonalizing the matrix produce state (MPS) transfer matrix in that sector yields an eigenvector $v$ that defines a non-local pair operator $\hat{\Delta}_v$ whose correlator decays slowest with distance. Consistent with zero-momentum pairing, the associated eigenvalue is found to be real. Correlating $\hat{\Delta}_v$ with a local pair
$\Delta_{ij}=\langle c^{\dagger}_i c^{\dagger}_j \hat\Delta_v\rangle$ (see Supplementary Material for details), and Fourier transform produces a momentum-resolved proxy gap function
$\tilde\Delta_{\mathbf k}\propto\langle c_{-\mathbf k}c_{\mathbf k}\rangle_v$~\cite{Stefan2024}.

To analyze our results in momentum space, we introduce magnetic Bloch states $\ket{\mathbf{k}}$, defined as simultaneous eigenstates of the magnetic translation operators $T_1$ and $T_2$. Due to the underlying magnetic algebra, we find that constructing these Bloch states so as to satisfy the point-group action $C_6 \ket{\mathbf{k}} = \ket{C_6 \mathbf{k}}$ on a cylindrical geometry imposes the unintuitive quantization condition $e^{i k_y L_y} = (-1)^{L_y/a}$ (see Supplementary Material for details). Consequently, for circumferences $L_y = 5a$ in Fig.~\ref{fig:pair}, the allowed $k_y$ values exclude the origin at $k_y = 0$.

Throughout the dome the phase of $\tilde\Delta_{\mathbf k}$ winds by $-6\pi$ around the hole-like $\Gamma$ pocket (Fig.~\ref{fig:pair}(c,d)), signaling chiral $f-if$ pairing. In the weak pairing limit, the Bogoliubov–de Gennes Chern number obeys $C_{\text{BdG}} = 2C + \ell$, where $C$ is the Chern number of the normal state Chern band~\cite{ReadGreen2000,QiHughesZhang2010,WangLianZhang2015} and $\ell$ is the winding around the \emph{hole} pocket. 
Taking $\ell = -3$ yields $C_{\text{BdG}} = -1$ and hence $c_- = -1/2$, so the superconducting phase hosts one clockwise-propagating Majorana edge mode. However, this computation assumes weak-pairing (e.g. near the SC-metal transition), 
and thus may not apply near the FCI-SC transition, a point we will return to. 

The proxy gap function evolves smoothly from the strong coupling side to the weak coupling side. The entire dome therefore corresponds to a chiral $f-if$ phase. On the FCI side the amplitude of $\tilde\Delta_{\mathbf k}$ is nearly uniform and the single-particle occupation $n_{\mathbf k}$ varies gently across the Brillouin zone, indicating tightly bound bosonic pairs—a BEC-like regime (Fig.~\ref{fig:pair}(a,c)). Closer to the metallic edge, $n_{\mathbf k}$ develops sharp steps at the non-interacting Fermi contour, and $\tilde\Delta_{\mathbf k}$ concentrates on these crossings (Fig.~\ref{fig:pair}(b,d)), hallmarks of weak-coupling, Fermi-surface pairing. On the $L_y = 5a$ and $7a$ cylinder the correlation lengths evolve smoothly across this range (Fig.~\ref{fig:pd}), indicating a continuous BEC–BCS crossover: tightly bound pairs on the insulating flank morph into weak-coupling Cooper pairs near the metal.
However, this cross-over need not extrapolate to the 2D limit, a point we will return to.

\begin{figure}[t]
 \centering
  \includegraphics[width=\columnwidth]{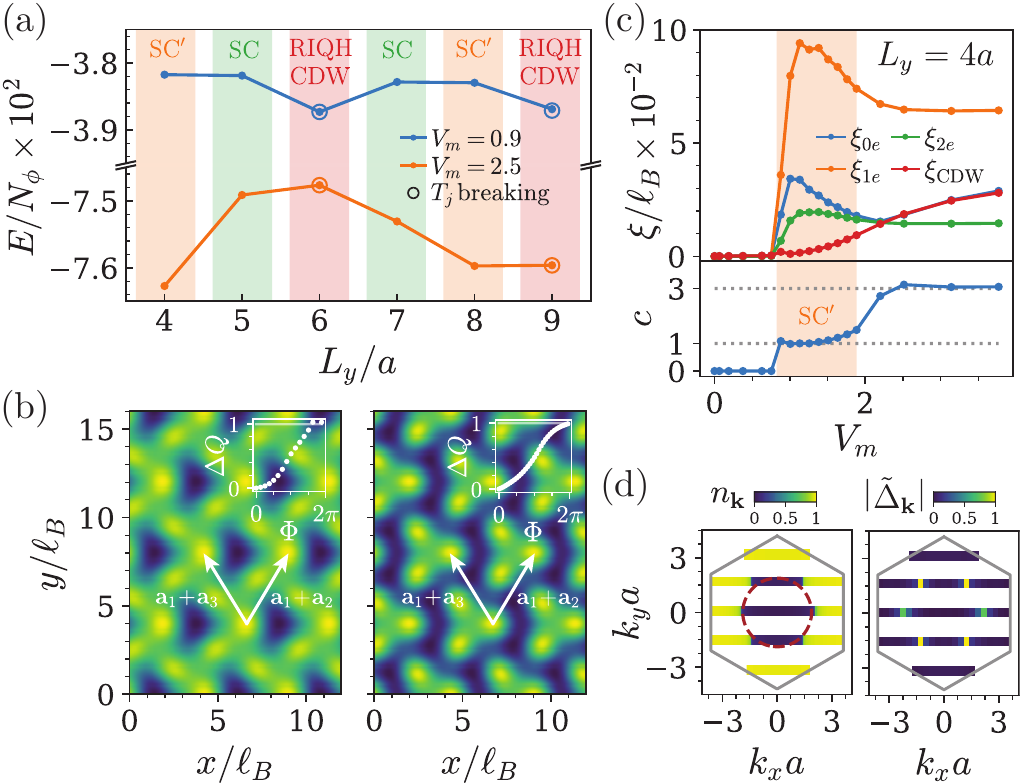}
\caption{\textbf{Competing instabilities.}
(a) Ground-state energy per flux quantum $E/N_\Phi$ versus cylinder circumference $L_y$ for two lattice amplitudes. Background shading identifies the phase inferred at each $L_y$: partially-gapped superconductor ($\text{SC}^\prime$, orange), superconductor (SC, green), and charge-density wave (CDW, RIQH, red). Open symbols denote runs in which translation symmetry is explicitly relaxed to favor a $\sqrt3\times\sqrt3$ CDW.
(b) Real-space filling factor $\nu_r(\mathbf r)$ of two closely competing $\sqrt3\times\sqrt3$ CDW obtained at $V_m=1.1$ on a $L_y=6a$ cylinder. White arrows indicate the enlarged Bravais vectors. Insets show average charge transferred $\Delta Q$ versus threaded flux $\Phi$. (c) Phase diagram for a $L_y = 4a$ cylinder.  
Top: correlation lengths in each quantum number sector.
Bottom: central charge $c$. The dotted lines are $c = 1$ and $3$ for visual guidance. 
The shaded window marks the partially-gapped superconductor $\text{SC}^\prime$ regime. (d) Single-particle occupation $n_{\mathbf k}$ and magnitude of the proxy gap function $\tilde{\Delta}_{\mathbf k}$ for the $\text{SC}^\prime$ obtained at $V_m=1.9$ on a $L_y=4a$ cylinder.}
\label{fig:CDW}
\end{figure}

\begin{figure}[t]
 \centering
  \includegraphics[width=\columnwidth]{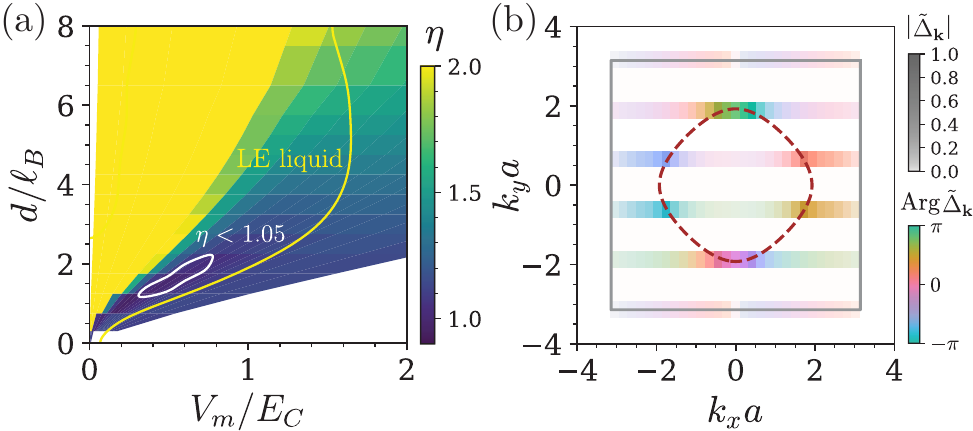}
\caption{\textbf{Robustness of the superconducting dome.}
(a) Luttinger exponent $\eta$ on a $L_y=5a$ cylinder as a function of lattice amplitude $V_m$ and gate distance $d$ for a screened Coulomb interaction. The yellow curve encloses the Luther–Emery (LE) liquid where $\xi_{2e}>\xi_{1e}$, while the inner white contour highlights the region with $\eta<1.05$.
(b) Momentum-resolved proxy gap function $\tilde\Delta_{\mathbf k}$ for a square-lattice potential at filling $\nu=11/15$ and $V_m=2.5$ on a $L_y=5a$ cylinder. The phase retains the $-6\pi$ winding similar to the triangular lattice case.}
\label{fig:square_d}
\end{figure}

\section{Competing instabilities and robustness of the superconductivity}

Within the triangular–lattice LLL model we find two phases compete closely with the chiral superconductor: (i) a commensurate $\sqrt3\times\sqrt3$ charge–density wave (CDW) with a re-entrant integer quantum Hall (RIQH) response (i.e. $\sigma_{xy} = e^2/h$ at $\nu = 2/3$) and (ii) a nearly gapless (or perhaps gapless) SC. We are able to explore the competition between these phases by varying the cylinder circumference: Table~\ref{tab:Ly} summarizes the ground state at each cylinder circumference explored, and Fig.~\ref{fig:CDW}(a) presents the corresponding energies.
\setlength{\tabcolsep}{5pt}
\begin{table}[h!]
\centering
\renewcommand{\arraystretch}{1.2}
\begin{tabular}{c|c|c}
\hline\hline
$L_y/a$ & Geometric feature & Ground state \\
\hline
$5,\;7$  & APBC                 & Superconductor (SC) \\
$4,\;8$  & PBC, solitary wire $k_y=0$       & Partially-gapped SC\\
$6,\;9$ & $\sqrt3\times\sqrt3$ commensurate & CDW (RIQH) \\
\hline\hline
\end{tabular}
\caption{\label{tab:Ly} Geometric effect that selects the ground state at different circumferences, examined inside the superconducting dome identified at $L_y=5a$.}
\end{table}

In contrast to $L_y = 5a$ and $7a$ where the robust superconductivity was observed, $L_y = 6a$ and $9a$ are commensurate with a $\sqrt{3} \times \sqrt{3}$ reconstruction and a charge-density wave (CDW) becomes favorable. DMRG actually resolves two such CDW states that are nearly degenerate with energy per flux differ by less than $10^{-5}$, yet they exhibit subtly different real-space textures as shown in Fig.~\ref{fig:CDW}(b). One resembles a hole Wigner crystal, with the depleted sites forming a triangular lattice; the other is closer to a Hall crystal~\cite{Tesanovic1989,Dong2024AHC1}, where the holes arrange into a honeycomb pattern. Despite their distinct appearances, both states share the same $\sqrt{3} \times \sqrt{3}$ periodicity and same quantized Hall response.

We diagnose the CDW order in two complementary ways. First, we run DMRG with the momentum quantum number conserved only modulo $3\times 2\pi/L_y$, allowing the MPS to break translation symmetry explicitly, which yields the the $\sqrt{3} \times \sqrt{3}$ modulation in Fig.~\ref{fig:CDW}(b). Alternatively, when we enforce full momentum conservation modulo $2\pi/L_y$, the ansatz form a cat state superposing the symmetry-related CDW patterns, and the corresponding finite-wave-vector correlation length $\xi_{\mathrm{CDW}}$ diverges, signaling long-range order without explicit symmetry breaking. From both approaches, the transition from the FCI to the CDW appears to be first order. Finally, by adiabatically threading flux through the cylinder \cite{zaletel2014flux} we verify that both CDW states possess Hall conductance $\sigma_{xy}=e^{2}/h$, i.e. they realize re-entrant integer quantum Hall phases (see insets of Fig.~\ref{fig:CDW}(b)).

As shown in Fig.~\ref{fig:CDW}(a), the energy per flux of the RIQH phase differs by less than $10^{-3}$ from the SC state at $L_y = 5a$ and $7a$.
The sign of this difference even appears to change with $V_m$, with $V_m = 0.9$ preferring RIQH and $V_m = 2.5$ preferring SC, indicating extremely close competition between the SC and RIQH phases. Future studies at larger systems sizes will be required to sort out their competition at fixed $V_m$.
Notably, RIQH states are experimentally observed on the electron-doped side of twisted MoTe$_2$~\cite{Xu2025FCISC} and in doped pentalayer graphene~\cite{Lu2024,Han2025,LuHanYao2025}, whereas superconductivity appears on the hole-doped side or when the moiré potential is weakened—mirroring the delicate balance we uncover numerically.

Finally, we consider $L_y = 4a$ and $8a$, whose behavior is peculiar.
Recall the momentum quantization condition is $e^{i k_y L_y} = (-1)^{L_y/a}$, and thus even / odd circumference effectively realize periodic / anti-periodic boundary conditions (PBC / APBC). Even $L_y/a$ is thus unique in that it accommodates $k_y = 0$ states which cut through the Fermi-surface (FS) as shown in Fig.~\ref{fig:CDW}(d). As for other $L_y$, we find evidence for three phases: the FCI, an intermediate phase we call SC', and a metal, with central charge $c = 0, 1, 3$ respectively at $L_y = 4 a$. Three $k_y$-modes cut through the FS, explaining $c = 3$ in the metal.

As shown in Fig.~\ref{fig:CDW}(c), the SC' state shows a SC correlation length $\xi_{2e}$ of the same order as the $L_y = 5a$ and $7a$ SC. However, the $\xi_{1e}$ correlation length is even longer, indicating near-gapless electron excitations. 
The origin of this behavior can be further understood from the proxy pairing function $\tilde{\Delta}_k$, which shows strong pairing at the $k_y \neq 0$ FS points, but weak pairing at the $k_y = 0$ FS (see Fig.~\ref{fig:CDW}(d)). 
As DMRG simulations at finite bond dimension can only lower-bound $\xi$, it is not clear whether the electron excitations remain truly gapless at $k_y = 0$ or merely have very large SC coherence length.
Since we are not aware of a selection rule that can prohibit pairing at $k_y = 0$, we suspect the latter.
The central charge of SC' is $c = 1$, also consistent with a superconductor.
The strong $k_y$ dependence of the pairing is possibly a quantitative artifact of thin cylinders; a useful check for future work will be to thread flux through the cylinder in order to realize APBC at $L_y = 4a$ and $8a$ to see if more uniform SC behavior returns. 

We next probe how a finite interaction range modifies the superconducting dome. Replacing the contact potential by a gated Coulomb interaction
\begin{equation}
  V(\mathbf{q})=E_C \frac{\tanh \left(q d\right)}{q \ell_B}
\end{equation}
interpolates smoothly from the $\delta$-function limit $d \to 0$ to the unscreened $1/r$ tail. 
Because the leading Haldane pseudopotential $V_{1}$ grows with $d$, the metal–FCI boundary shifts to larger bandwidth $V_{m}/E_{C}$. Yet the qualitative structure survives: a superconducting dome still interposes between the FCI and the metal, and its interior remains a Luther–Emery (LE) liquid with algebraic pair correlations (see Supplementary Materials). As shown in Fig.~\ref{fig:square_d}(a), The LE liquid behavior survives $d \lesssim 10\ell_{B}$ and the Luttinger exponent $\eta$ touches unity for $d \lesssim 2\ell_{B}$ on a $L_{y}=5a$ cylinder.

This window matches experimental length scales. 
For a $3.8^{\circ}$ twisted MoTe$_2$ bilayer, the moiré period $a_{\mathrm M}\simeq5.2$~nm yields effective $\ell_{B}\approx1.9$~nm; hBN spacers of $10 - 30$~nm correspond to $d/\ell_{B}\approx5 - 15$, overlapping the LE liquid regime~\cite{Xu2025FCISC}. 
In hBN-aligned rhombohedral pentalayer graphene, $a_{\mathrm M}\simeq14$~nm gives $\ell_{B}\approx5.2$~nm; gate distances of $20 - 40$~nm translate to $d/\ell_{B}\approx4 - 7$, squarely inside the superconducting regime~\cite{Han2025}.

Chiral pairing also survives a drastic change of lattice geometry. 
Replacing the triangular lattice by a square lattice leaves only a faint LE liquid window at the parent filling $\nu=2/3$, but a modest electron doping to $\nu=11/15$ dramatically enlarges the window (see Supplementary material). The momentum-resolved gap shown in Fig.~\ref{fig:square_d}(b) still exhibits a $-6\pi$ phase winding, confirming that the $f-if$ order and the underlying mechanism are properties of the FCI melting itself rather than of any Fermi-surface details. Taken together, the gated-Coulomb map and the square-lattice test establish that the chiral superconductivity we find is remarkably robust—persisting across realistic screening lengths, carrier densities, and even disparate lattice geometries.

\section{Discussion}
Our numerical calculations demonstrate that a fully spin-polarized, repulsive electron fluid can pass directly from a fractional Chern insulator to a chiral superconductor at fixed filling. One possible interpretation of this result is through a parton construction $c=b f$. We place the bosonic parton $b$ in a bosonic IQH state with Hall conductance $\sigma^{b}_{xy}=2e^{2}/h$.
At $\nu=2/3$ Galilean invariance then forces the fermionic parton $f$ to a $C_{f}=1$ Landau level, reproducing the Jain FCI (see Suplementary material). More generally, the low-energy theory is captured by the Chern-Simons Lagrangian
\begin{equation}
\mathcal{L}=\frac{2}{4 \pi} (A-a) d(A-a)+\frac{C_f}{4 \pi} a d a
\end{equation}
where $a d b \equiv \epsilon^{\mu\nu\lambda} a_\mu \partial_\nu b_\lambda$, $C_f$ is the Chern number of the $f$-parton band, and $a$ is the emergent gauge field that binds the partons into the physical electron. Bandwidth tuning can instead invert the $f$ band to Chern number $C_{f}=-2$. The bosonic and fermionic Chern–Simons terms then cancel, the emergent gauge field becomes gapless~\cite{MaissamLaughlin}, resulting in a chiral superconductor with edge chiral central charge $c_-=-2$, obtained without external doping. It lands in the same topological class as the anyon superconductor proposed in Ref.~\cite{Darius2024} and has similarities in spirit, but here arises from a simple band inversion rather than from separate Hall liquids of dopant. The accompanying critical theory is the QED$_3$-Chern-Simons fixed point that also governs bosonic Laughlin-to-superfluid transitions~\cite{MaissamLaughlin,TaigeTransition,JYLQED}.

Magnetic translations at filling $\nu = 2/3$ force the Chern number $C_f$ to change in multiples of three, so the $C_f=-2$ route naturally selects superconductivity unless the lattice enlarges. A spontaneous $\sqrt3\times\sqrt3$ reconstruction releases this constraint. Choosing $C_f=+2$ produces the re-entrant IQH CDW seen both in our numerics and experiments~\cite{Xu2025FCISC,Lu2024,LuHanYao2025}; in the parton theory this CDW coexists with a neutral $U(1)_{-4}$ topological order, although our present DMRG data cannot yet resolve its presence. Setting $C_f = 0$ instead yields a CDW without Hall response but coexist with a neutral semion topological order—a state discussed in Ref.~\onlinecite{XueYangSenthil} and consistent with the phase observed at large displacement field in twisted MoTe$_2$~\cite{Cai2023,Park2023}. Thus the parton picture unifies the superconductor, the RIQH CDW, and the $\sigma_{xy}=0$ CDW as proximate descendants of the same FCI, while a direct FCI-metal transition lacks any field theory description—possibly explaining why our numerics and the experiments always find an intervening ordered phase.

Whether the $ f - if$ condensate we observe truly coincides with the parton-derived phase remains unclear. 
In the weak-coupling limit, our weak-coupling $ f - if$ state carries edge chiral central charge $c_- =-1/2$, whereas the parton construction yields $c_-=-2$; the two phases differ by three chiral $p + ip$ Majorana modes. 
More broadly, spin-polarized superconductors are expected to have half-integer $c_-$ in the BCS limit and integer $c_-$ in the BEC limit, apparently precluding a smooth crossover~\cite{GurarieRadzihovsky2007}. 
In the 2D limit, there are two possible resolutions: (i) the $c_-=-2$ parton superconductor sheds its extra Majorana before reaching the metal; (ii) the superconducting dome is adiabatically connected to the weak-coupling $f - if$ phase, but ends through a first-order phase transition into the FCI phase. 
Yet our \emph{cylinder} numerics at $L_y = 5a$ and $7a$ appear to have a continuous BEC-to-BCS evolution.
Such a crossover could be a 1D-artifact from the $k_y$ momentum quantization on a cylinder, so that the momentum points where the gap closes at the BEC-BCS transitions are avoided. 
If the closing occurs at the $\Gamma$ point, this could then be the origin of the large correlation length observed for PBC cylinders $L_y = 4a$ and $8a$.
A direct numerical work will be required to decide between these scenarios.

A complementary, weak-coupling interpretation is that interactions near the FCI become effectively attractive at intermediate length scales~\cite{Kim2025TCS}. 
Very recent numerical calculations show that Laughlin anyons can themselves bind into charge-$2e/3$ composites and even Cooper pairs~\cite{Xu2025Clusters,GattuJain2025,TaigePairing}. 
Once the Laughlin gap softens, these Cooper pairs may condense into the chiral superconductor we observe, providing a microscopic route fully compatible with weak-coupling expectations. Regardless of whether the superconductor emerges through a continuous FCI-to-SC transition or via an effective mid-range attraction among anyons, the bandwidth-driven instability uncovered here offers a generic, dopant-free pathway to spin-polarized chiral pairing and naturally accounts for its close rivalry with re-entrant Hall order in current moiré heterostructures.

Taken literally, these results translate into various experimental expectations. 
Because the superconducting dome nucleates precisely where the topological gap first closes, increasing the twist angle—and hence the ratio of moiré potential to Coulomb energy—should broaden the dome so that robust chiral pairing already appears at the bare filling $\nu = 2/3$, without the need for extra carriers or deliberate suppression of the moiré potential. 
Conversely, slight doping or any perturbation that weakens the periodic potential relaxes the commensuration condition that stabilizes the $\sqrt3\times\sqrt3$ charge–density wave, thereby tipping the balance toward superconductivity. 
A systematic sweep of twist angle and carrier density in twisted MoTe$_2$ and rhombohedral pentalayer graphene can therefore test the ``melt-the-FCI'' mechanism and controllably navigate between the superconducting and re-entrant integer quantum Hall (RIQH) regimes of a spin-polarized Chern band.

A perpendicular magnetic field offers an orthogonal tuning knob. 
Once the field is strong enough to quench the superconducting condensate, the system is predicted to enter the competing RIQH phase, since superconductivity is inherently sensitive to out-of-plane fields, whereas the orbital moment associated with the Hall response favors the RIQH state. This field-induced crossover has already been reported in twisted MoTe$_2$ devices~\cite{Xu2025FCISC}. 
Inside the RIQH regime, real-space probes such as scanning tunnelling microscopy (STM) should detect the anticipated $\sqrt3\times\sqrt3$ density modulation~\cite{jiang2019charge,kerelsky2019maximized,li2021generalized}, providing a direct structural fingerprint of the charge-density wave (CDW) and a decisive test of the microscopic picture developed here.
 
\emph{Note added--} During writing of this manuscript, we become aware of Ref.~\onlinecite{ShiSenthil2025Anyon,Zhang2025Holon,Pichler2025}, which discuss a similar field-theory construction of bandwidth-tuned FCI to superconductor transition.

\textbf{Acknowledgments.} We are especially grateful to Ya-Hui Zhang for helpful discussions. We thank Zhihuan Dong for collaboration on a related project. We acknowledge Ashvin Vishwanath, Daniel Parker, Dung-Hai Lee, T. Senthil, Tianle Wang, Sian Yang for discussion. This work is supported by the Simons Collaboration on Ultra-Quantum Matter, which is a grant from the Simons Foundation (1151944, MZ). This research uses the Lawrencium computational cluster provided by the Lawrence Berkeley National Laboratory (Supported by the U.S. Department of Energy, Office of Basic Energy Sciences under Contract No. DE-AC02-05-CH11231).

\bibliography{arxiv}






 


\end{document}